\documentclass[10pt,twocolumn,pra,showpacs,floatfix]{revtex4}
\usepackage{dcolumn}                 
\newcolumntype{w}[1]{D{.}{.}{#1}}
\newcommand*{\centt}[1]{\multicolumn{1}{c}{#1}}
\newcommand*{\cent}[1]{\multicolumn{1}{c}{$#1$}}
\usepackage{times}
\usepackage{nicefrac}
\usepackage{amsmath}
\usepackage{amsfonts}
\usepackage{amssymb}
\usepackage{amsthm}

\begin{document}
\preprint{Version 1.0}
\title{D1 and D2 lines in  $^{6,7}$Li including QED effects}

\author{M. Puchalski}

\affiliation{Faculty of Chemistry, Adam Mickiewicz University, 
             Grunwaldzka 6,60-780 Pozna\'n, Poland }

\author{D. K\c edziera}

\affiliation{Faculty of Chemistry, Nicolaus Copernicus University, 
             Gagarina 7, 87-100 Toru\'n, Poland}

\author{K. Pachucki}

\affiliation{Faculty of Physics, University of Warsaw,
             Ho\.{z}a 69, 00-681 Warsaw, Poland}

\begin{abstract}
Accurate theoretical predictions including leading QED corrections for 
$2^2P_{1/2} - 2^2S_{1/2}$ (D1) and  $2^2P_{3/2} - 2^2S_{1/2}$ (D2) transition energies
have been obtained for  $^{6,7}$Li isotopes. Our results for the Bethe logarithms 
$\ln k_0(2^2S) = 5.178\,169(4)$, and for mass polarization corrections 
$\Delta \ln k_0(2^2S) = 0.113\,81(3)$ are in disagreement with ones obtained recently in the literature.
In contrast, results $\ln k_0(2^2P) = 5.179\,81(7)$ and 
$\Delta \ln k_0(2^2P) = 0.111\,3(5)$ are in good agreement with them. 
From our theoretical predictions and recent measurements of  $^{6,7}$Li D-lines at NIST,
we determine the mean square charge radius difference between $^7$Li and $^6$Li nuclei,
in agreement with determinations based on the $3^2S_{1/2}-2^2S_{1/2}$ transition, what 
demonstrates consistency of atomic spectroscopy determination of fundamental properties of nuclei. 
\end{abstract}
     
\maketitle
\section{Introduction}
Accurate spectroscopy of the lithium atom allows one to test computational approaches
to many electron systems including quantum electrodynamic (QED) effects.
Comparison with measured values not only verifies theoretical predictions,
but also experimental values as many discrepancies have been reported in the literature.  
A very good example is the recent measurement of D-lines performed at NIST \cite{sansonetti2011, brown2012},
which was stimulated by a long standing discrepancy among theoretical and experimental results
for the fine structure splitting $2 P_{3/2} - 2 P_{1/2}$, and its difference
between isotopes. This has been calculated in many works and only
recently, it was discovered a significant effect of the hyperfine mixing of $P$-levels \cite{lit_fine}.
From the experimental side, the presence of the nearby level, whose natural width overlaps
with measured transitions, may contribute significantly to the lineshape and to the linewidth 
\cite{hessels2011, sansonetti2011}. When it is taken into account in determination of line positions,
the most recent experimental result \cite{sansonetti2011,brown2012} becomes in a very good agreement 
with theoretical predictions.
Moreover, assuming that theoretical predictions and recent measurements \cite{brown2012} are 
as accurate as claimed, the mean square nuclear charge radii difference between $^6$Li
and $^7$Li can be determined. Our result and that presented in experimental work \cite{brown2012},
are in an agreement with the one obtained from $3 S-2 S$ transition. This strongly supports
the atomic spectroscopy determination of nuclear charge radii, not only for stable nuclei,
but also for very short lived $^{11}$Li and $^{11}$Be isotopes, which can not be studied by other means.

In this work we present the most accurate calculation of relativistic and leading 
quantum electrodynamics corrections to lithium D-lines. The calculational approach 
is similar to that developed previously by Drake and Yan in Ref.
\cite{yan_rel}, with some differences regarding
evaluation of three electron integrals. We use recurrence relations,
derived in \cite{remiddi}. They are sufficiently stable and fast
to obtain nonrelativistic energy levels to $10^{-12}$ precision in a one day
calculations \cite{lit_wave}. Apart from nonrelativistic energy,
other contributions to energy, relativistic and QED ones require 
evaluation of more complicated matrix elements,
what is probably, the most challenging part of these calculations. Here, we
use an approach developed in \cite{remiddi, lit_rel}, and obtain results in good agreement
with that by Drake and Yan in \cite{yan_rel}, 
with exception of Bethe logarithms and it's mass polarization corrections.
They have not so far, been verified with competitive accuracy, and indeed our result for $2S$ state
is not in an agreement with that obtained in \cite{yan_rel}.

\section{Expansion of the energy levels}
We follow the approach used in the former calculations on the relativistic and QED 
effects for three-electron systems \cite{lit_rel,lit_fine}. Here we
concentrate on D1 and D2 transition energies and the related isotope mass shifts. 

Energy levels are expanded in a power series of the fine structure constant 
$\alpha$ and the reduced electron mass to nuclear mass ratio $\eta=-\mu/m_N$
\begin{eqnarray}
E &=& m\,\alpha^2\,\bigl[{\cal E}^{(2,0)}+\eta\,{\cal E}^{(2,1)}+\eta^2\,{\cal E}^{(2,2)}\bigr] \nonumber \\
&& +\,m\,\alpha^4\,\bigl[{\cal E}^{(4,0)}+{\cal F}^{(4,0)}+\eta\,({\cal E}^{(4,1)}+{\cal F}^{(4,1)} )\bigr] \nonumber \\
&& +\,m\,\alpha^5\,\bigl[{\cal E}^{(5,0)}+\eta\,{\cal E}^{(5,1)}\bigr] \\
&& +\,m\,\alpha^6\,{\cal E}^{(6,0)} +m\,\alpha^7\,{\cal E}^{(7,0)} + \ldots,
\label{Eexpansion}
\end{eqnarray}
with dimensionless spin independent (centroid) ${\cal E}^{(m,n)}$ 
and spin dependent ${\cal F}^{(m,n)}$ coefficients. 
All neglected terms denoted by ``$\ldots$'' involve higher powers of $\alpha$ and
the mass ratio  $\eta$ and thus are small in comparison  to the accuracy of the final
results. The leading contribution ${\cal E}^{(2,0)}\equiv {\cal E}_0$ is a
solution of the Shr{\" o}dinger equation with the clamped nucleus
\begin{eqnarray}
{\cal H}_0 \Psi &=&  {\cal E}_0 \, \Psi \,, \\ 
{\cal H}_0 &=& \sum_a \bigg[ \frac{p_a^2}{2} - \frac{Z}{r_a} \biggr] +\sum_{a<b} \frac{1}{r_{ab}}. \label{h0}
\end{eqnarray}
The trial wave function $\Psi$ is a linear combination of $\psi$ elements, the antisymmetrized product ${\cal A}$
of the spatial function $\phi$ and the spin function $\chi$ \cite{lit_wave,lit_rel}
\begin{eqnarray}
\psi_S &=& {\cal A}[\phi_S\,\chi]\,,
\label{psiS}\\
\vec \psi_{Pa} &=& {\cal A}[\vec \phi_{Pa}\,\chi]\,,
\label{psiP}\\
\phi_S &=& r_{23}^{n_1}\,r_{31}^{n_2}\,r_{12}^{n_3}\,r_{1}^{n_4}\,r_{2}^{n_5}\,r_{3}^{n_6}\,e^{-w_1\,r_1-w_2\,r_2-w_3\,r_3},
\label{phiS} \\
\vec \phi_{Pa} &=& \vec r_a\,\phi_S
\label{phiP}\\
\chi &=& (\alpha(1)\,\beta(2)-\beta(1)\,\alpha(2))\,\alpha(3)\,,
\label{14}
\end{eqnarray}
with $n_i$ being non-negative integers, $w_i\in {R}_+$, and the subscript $a=1,2,3$.
The matrix element of the  ${\cal H}_0$ in Eq. (\ref{h0})
(or of any spin independent operator) can be expressed after eliminating spin part $\chi$ as
\begin{eqnarray}
\langle\psi|H_0|\psi'\rangle &=& \langle
2\,\phi(1,2,3)+2\,\phi(2,1,3)-\phi(3,1,2)\nonumber \\ 
&& \hspace{-2.2cm}-\phi(2,3,1)-\phi(1,3,2)-\phi(3,2,1)|
 H_0\,|\phi'(1,2,3)\rangle\,.
\end{eqnarray}

In comparison to our former calculations in Ref. \cite{lit_rel}, we have improved in
\cite{slater1} numerical results for nonrelativistic energy of the ground
state, 
\begin{equation}
{\cal E}_0(2 S) = -7.478\,060\,323\,910\,2(2) 
\end{equation}
and here for the first excited state
\begin{equation}
{\cal E}_0(2 P) = -7.410\,156\,532\,651\,4(9).
\end{equation}
They are however, about $2$   
orders of magnitude less accurate, compared to the most recent large scale
calculations by Wang {\it et al.} \cite{Wang_2012}. 
We note, that overall theoretical predictions are not limited, at present,
by nonrelativistic energy, but by the unknown $m\,\alpha^6$ contribution.

In the infinite nuclear mass limit, we neglect in the nonrelativistic Hamiltonian
the mass polarization term, which is included later as the perturbation to ${\cal H}_0$
\begin{eqnarray}
{\cal H}_{\rm mp} &=& -\eta \sum_{a<b} \vec p_a \cdot \vec p_b \,.  
\label{hmp}
\end{eqnarray}
The first order perturbation to the wave function 
$\Psi \rightarrow \Psi + \eta \, \delta\Psi_{\rm mp}$ is defined by
\begin{eqnarray}
\delta \Psi_{\rm mp} &=& -\frac{1}{{\cal E}_0-{\cal H}_0} \, \sum_{a<b} \vec p_a \cdot \vec p_b \, \Psi\,.
\label{psimp}
\end{eqnarray}
Lets us denote $\langle \ldots \rangle_{\rm mp} = 2 \langle \Psi | \ldots |
\delta \Psi_{\rm mp} \rangle$ 
as a correction originating from the mass polarization perturbation of the wave function. Then, the first and 
the second order finite mass nonrelativistic corrections are obtained from the mass scaling and the mass 
polarization corrections
\begin{eqnarray}
{\cal E}^{(2,1)} &=& {\cal E}^{(2,0)} - \sum_{a<b}\,\langle \vec p_a \cdot \vec p_b \rangle,  \\
{\cal E}^{(2,2)} &=& -\sum_{a<b}\,\biggl[ \langle \vec p_a \cdot \vec p_b \rangle + \frac{1}{2} \, 
 \langle \vec p_a \cdot \vec p_b\rangle_{\rm mp} \biggr]\,,
\end{eqnarray}
%
%

%
%

From the Breit-Pauli Hamiltonian, we extract the spin independent part
\begin{eqnarray}
\label{h40}
{\cal H}^{(4,0)} &=& \sum_a \biggl[ -\frac{\vec p^{\,4}_a}{8}  + \frac{ \pi\,Z\,\alpha}{2}\,\delta^3(r_a)\biggr ]
 \\
&&  +\sum_{a<b} \biggl [\pi\, \delta^3(r_{ab})
-\frac{1}{2}\, p_a^i\,\biggl(\frac{\delta^{ij}}{r_{ab}}+\frac{r^i_{ab}\,r^j_{ab}}{r^3_{ab}}
\biggr)\, p_b^j \biggr].\nonumber
\end{eqnarray}
leading to the coefficient ${\cal E}^{(4,0)} = \langle {\cal H}^{(4,0)}\rangle$. The finite 
nuclear mass correction ${\cal E}^{(4,1)}$ consists of the mass scaling term ${\cal E}^{(4,1)}_{\rm ms}$,
the mass polarization perturbation  $\langle {\cal H}^{(4,0)}\rangle_{\rm mp}$,
and the explicit recoil term ${\cal H}^{(4,1)}_{\rm r}$ coming from 
the Breit interaction between electrons and the nucleus,
\begin{eqnarray}
{\cal E}^{(4,1)} &=& {\cal E}^{(4,1)}_{\rm ms} + \langle {\cal H}^{(4,0)} \rangle_{\rm mp} + \langle {\cal H}^{(4,1)}_{\rm r} \rangle, \\
{\cal E}^{(4,1)}_{\rm ms} &=& -4 \sum_a \frac{1}{8}\langle \vec p^{\,4}_a \rangle
+ 3 \, \biggr[  \frac{ \pi\,Z\,\alpha}{2}\,\sum_a\,\langle \delta^3(r_a) \rangle  \\
&&
\hspace{-1cm} + \pi\,\sum_{a<b}\, \langle \delta^3(r_{ab}) \rangle
-\frac{1}{2}\,\sum_{a<b}\, \biggl \langle p_a^i\,\biggl(\frac{\delta^{ij}}{r_{ab}}+\frac{r^i_{ab}\,r^j_{ab}}{r^3_{ab}}
\biggr)\, p_b^j \biggr \rangle \biggr],\nonumber\\
{\cal H}^{(4,1)}_{\rm r} &=& - \frac{Z}{2} \,
\,\sum_{a,b} p_a^i\,\biggl(\frac{\delta^{ij}}{r_{a}}+\frac{r^i_{a}\,r^j_{a}}{r^3_{a}}
\biggr)\, p_b^j.
\end{eqnarray}
The spin-orbit coupling is given by
\begin{eqnarray}
{\cal H}^{(4,0)}_{\rm fs} &=& \sum_a \frac{Z}{2\,r_a^3}\,\vec s_a\,
(g-1)\,\vec r_a\times\vec p_a 
 \\ &&+
\sum_{a\neq b}\frac{1}{2\,r_{ab}^3}\,
\vec s_a\bigl[g\,\vec r_{ab}\times\vec p_b
-(g-1)\,\vec r_{ab}\times\vec p_a\bigr]\nonumber \\
{\cal F}^{(4,0)} &=& \langle {\cal H}^{(4,0)}_{\rm fs}\rangle \\
{\cal H}^{(4,1)}_{\rm fs,r} &=& -\sum_{a,b} \frac{Z}{2\,r_a^3}\,\vec s_a\,g\,\vec r_a\times\vec p_b \\
{\cal F}^{(4,1)} &=&  3\,\langle {\cal H}^{(4,0)}_{\rm fs}\rangle + 
\langle {\cal H}^{(4,0)}_{\rm fs} \rangle_{\rm mp} + \langle {\cal H}^{(4,0)}_{\rm fs,r}\rangle\,.
\end{eqnarray}
with spin matrices represented using Pauli matrices $\vec s_a = \vec \sigma_a/2$. The matrix element of spin-orbit terms are obtained for $n P_{1/2}$ state by using
\begin{eqnarray}
\langle \vec \psi_a|\sum_{c=1}^3\vec Q_c\cdot\vec\sigma_c| \vec \psi_b\rangle &=&
i\,\Bigl\langle \vec\phi_a(1,2,3) \Bigl| \\
&& \hspace{-4cm} -2\,\vec Q_3\times\bigl[\vec\phi_b(1,2,3)+\vec\phi_b(2,1,3)\bigr] \nonumber\\
&& \hspace{-4cm} +(\vec Q_1-\vec Q_2+\vec Q_3)\times\bigl[\vec\phi_b(2,3,1)+\vec\phi_b(3,2,1)\bigr]\nonumber\\
&& \hspace{-4cm} +(-\vec Q_1+\vec Q_2+\vec Q_3)\times\bigl[\vec\phi_b(1,3,2)+\vec\phi_b(3,1,2r)\bigr]\Bigr\rangle. \nonumber
\end{eqnarray}
and for the $n P_{3/2}$ state ${\cal F}(P_{3/2}) = -1/2\,{\cal F}(P_{1/2})$.

%
%

Leading QED correction of order $m\,\alpha^5$ is given by
\begin{eqnarray}
{\cal E}^{(5,0)} &=& \frac{4\,Z}{3}\,\left[\frac{19}{30}+\ln(\alpha^{-2}) - \ln k_0\right]\,
\sum_a\,\langle \delta^3(r_a) \rangle \nonumber \\ && +
 \left[\frac{164}{15}+\frac{14}{3}\,\ln\alpha
\right]\,\sum_{a<b} \,\langle \delta^3(r_{ab}) \rangle \nonumber \\ &&
-\frac{7}{6\,\pi} \,\sum_{a<b} \,\biggl \langle P\left(\frac{1}{r_{ab}^3} \right) \biggr \rangle ,
\label{e50}
\end{eqnarray}
where the Bethe logarithm $\ln k_0$ has the form
\begin{eqnarray}
\ln k_0 &\equiv&  \nonumber \\ && \hspace{-1cm} \frac{\left\langle \sum_a\vec{p}_a \,(H_0-E_0)\,
\ln\bigl[2\,(H_0-E_0)]\,\sum_b\vec{p}_b \right\rangle}{2\,\pi\,Z\,
\sum_c \langle \delta^3(r_c) \rangle} 
\label{Bethelog}
\end{eqnarray}
and the Araki-Sucher term
\begin{eqnarray}
\langle\phi| P\left(\frac{1}{r^3}\right)|\psi\rangle &=& \lim_{a\rightarrow 0}\int {\rm d}^3 r\,\phi^*(\vec r)\biggl[  \label{Prabm3}\\
&&  \frac{1}{r^3}\,\Theta(r-a) + 4\,\pi\,\delta^3(r)\, (\gamma+\ln a) \biggr]\,\psi(\vec r). \nonumber
\end{eqnarray}

The finite nucleus mass correction to the leading QED contribution has the form
\begin{eqnarray}
{\cal E}^{(5,1)} &=& {\cal E}^{(5,1)}_{\rm ms}  + {\cal E}^{(5,1)}_{\rm mp} + {\cal E}^{(5,1)}_{\rm r}.
\label{e51}
\end{eqnarray}
The mass scaling part is given by
\begin{eqnarray}
{\cal E}^{(5,1)}_{\rm ms} &=& 3\,{\cal E}^{(5,0)} - \frac{4\,Z}{3}\,\sum_a\,\langle \delta^3(r_a) \rangle 
+ \frac{14}{3}\,\sum_{a<b}\,\langle \delta^3(r_{ab}) \rangle \, ,  \nonumber \\
\label{e51ms}
\end{eqnarray}
Apart from the scaling factor three for all operators in Eq. \eqref{e50}, there are
additional two terms from the mass scaling of the Bethe logarithm and
$P(1/r^3)$ operator. The direct recoil term ${\cal E}^{(5,1)}_{\rm r}$ is 
known as the Salpeter correction
\begin{eqnarray}
{\cal E}^{(5,1)}_{\rm r} &=& -\frac{Z^2}{3} \biggr [\frac{62}{3} + \ln(\alpha^{-2}) - 8\,\ln k_0 \biggl]\, \sum_a\, \langle \delta^3(r_a) \rangle  \nonumber \\
&& +Z^2\,\frac{7}{6\,\pi} \,\sum_a\, \biggl \langle P\left(\frac{1}{r_{a}^3} \right) \biggr \rangle,
\label{e51r}
\end{eqnarray}
and ${\cal E}^{(5,1)}_{\rm mp}$ is due to the mass polarization correction 
to expectation value of the operators in Eq.~\eqref{e50}. 

%
%

The relativistic $m\,\alpha^6$ corrections for few-electron atoms are very
difficult to calculate, and closed formula has been obtained only for
two-electron systems. The result for three- and more-electron systems
\cite{fw} contains divergences, which elimination has not yet been performed.
For three-electron systems we use an approximate formula on the basis of hydrogenic values including
dominating electron-nucleus one-loop radiative correction \cite{eides}
\begin{eqnarray}
{\cal E}^{(6,0)} &=& Z^2\,\pi\,\left[\frac{427}{96} - 2 \ln(2) \right] \,\sum_a \langle\delta^3(r_a) \rangle\,.
\label{e60}
\end{eqnarray}
We neglect electron-electron radiative corrections and the purely relativistic corrections, as we expect them to be
relatively small, of order 10\%. This is the leading source of uncertainty in the theoretical predictions for transition frequencies. 
Similarly, the finite nuclear mass correction, which is significant for the isotope mass shift,
is also estimated on the basis of the known hydrogenic values. 
Apart from the mass scaling and the mass polarization corrections to 
the Eq.~\eqref{e60},  the radiative recoil and pure recoil corrections have been included
as folows
\begin{eqnarray}
{\cal E}^{(6,1)} &=& 3\,{\cal E}^{(6,0)} + {\cal E}^{(6,0)}_{\rm mp} \nonumber\\
&&	 - \pi\,\biggl\{Z^2\,
            \left[\frac{35}{36} - \frac{448}{27 \pi^2} - 2 \ln(2)
            +\frac{6 \zeta(3)}{\pi^2} \right] \nonumber \\ 
&&          + Z^3 \,
            \left[4 \ln(2) - \frac{7}{2} \right]\biggr\}\,\sum_a \langle\delta^3(r_a) \rangle
\end{eqnarray}
Relativistic and QED contributions to the fine structure have been described
in details in Ref. \cite{lit_fine}, and we note that the hyperfine interaction
leads to a significant shift in the $^7$Li fine structure. 

Due to the numerical significance for transition energies, one
estimates the $m\,\alpha^7$ contribution on the basis of formulas for hydrogenic systems \cite{eides},
\begin{eqnarray}
{\cal E}^{(7,0)}_{\rm H}(n) &=& m\,\frac{\alpha}{\pi}\,\frac{(Z\,\alpha)^6}{n^3}\,
              \bigl[A_{60}(n)+\ln(Z\,\alpha)^{-2}\,A_{61}(n) \nonumber \\
             && +\ln^2(Z\,\alpha)^{-2}\,A_{62}\bigr]
              +m\,\Bigl(\frac{\alpha}{\pi}\Bigr)^2\,
               \frac{(Z\,\alpha)^5}{n^3}\,B_{50}\nonumber \\ &&
              +m\,\Bigl(\frac{\alpha}{\pi}\Bigr)^3\,
              \frac{(Z\,\alpha)^4}{n^3}\,C_{40}.
\end{eqnarray}
It involves one-, two-, and three-loop corrections, and values of $A,B$, and
$C$ coefficients may be found in \cite{eides}.
Following Ref. \cite{drake_book} these hydrogenic values of order $m\,\alpha^7$ are
extrapolated to lithium, according to
\begin{eqnarray}
{\cal E}^{(7,0)}(Z) &=& \bigl[2\,{\cal E}^{(7)}(1S,Z) +{\cal E}^{(7)}(nX,Z-2)\bigr] \nonumber \\
&&  \times \,\frac{\langle \delta^3(r_1) + \delta^3(r_2) + \delta^3(r_3)\rangle_{\rm Li}}
{2\,\langle \delta^3(r)\rangle_{1S,Z}
+ \langle\delta^3(r)\rangle_{nX,Z-2}},
\end{eqnarray}
for $X=S$, and for states with higher angular momenta ${\cal E}^{(7)}(nX,Z)$ is neglected.
We assume this approximate formula to be accurate to 25\%.

%
%

Beyond QED, there are corrections due to the finite nuclear size. 
The leading order correction $m \alpha^4$ is given by
\begin{eqnarray}
{\cal E}_{\rm fs}^{(4,0)} &=& \frac{2 \pi\,Z}{3}\,\frac{r_c^2}{\lambdabar^2} 
\,\sum_a  \langle \delta^3(r_a) \rangle,
\label{c40}
\end{eqnarray}
where $r_{\rm c}^2$ is the averaged square of the charge radius and
$\lambdabar$ is the electron Compton wavelength divided by $2\,\pi$. 
We additionally include a logarithmic relativistic correction to the wave function at the origin
\begin{equation}
{\cal E}_{\rm fs,log}^{(6,0)} = -(Z\,\alpha)^2\,\ln(Z\,\alpha\,m\,r_{\rm c})\,{\cal E}^{(4,0)}_{\rm fs}\,.
\label{clog60}
\end{equation}
and the finite mass correction ${\cal E}_{\rm fs}^{(4,1)}$ which consists of 
the mass scaling and mass polarization correction to the Eq.~\eqref{c40}.
One determines nuclear charge radii
from the difference between experimental and theoretical isotope shift by using
\begin{eqnarray}
\Delta\nu_{\rm exp}-\Delta\nu_{\rm the} &=& C_{AB}\,(r_{{\rm c}\,A}^2-r_{{\rm c}\,B}^2)\,, \\
C_{AB} &=& \frac{{\cal E}^A_{\rm fs} - {\cal E}^B_{\rm fs}}{r_{{\rm c}\,A}^2-r_{{\rm c}\,B}^2}\,.
\end{eqnarray}
the described finite nuclar size ${\cal E}_{\rm fs}$
for isotopes $A$ and $B$.

\section{Bethe logarithm and its mass polarization correction}

Bethe logarithm is the most demanding term 
in accurate numerical evaluation among all operators in Eq.~\eqref{Eexpansion}.  
In this work we use the integral representation introduced by Schwartz
\cite{schwartz_blog} with compact set of formulas \cite{ligauss} given by 
\begin{eqnarray}
\ln k_0 &=& \frac{1}{D} \int_0^1 dt \, \frac{f(t)-f_0-f_2t^2}{t^3} 
\label{intlnk0}\\
f(t) &=& - \omega \,\bigg \langle \vec P \frac{1}{{\cal E}_0-{\cal H}_0-\omega} \vec P \bigg \rangle\,,
\label{ft}\\
t &=& \frac{1}{\sqrt{1+2\,\omega}}\\
\vec P &=& \sum_a \vec p_a\,, \qquad D = 2 \pi Z  \sum_a \langle \delta^3(r_a) \rangle\,, \\
f_0 &=&  \langle \vec P^2 \rangle\,, \qquad f_2 = -2 D\,, 
\end{eqnarray}
The integrand in Eq.~\eqref{intlnk0} is a regular function with a reasonable good
numerical convergence calculated at 100 equally spaced 
points in $t \in [0,1]$ variable. The critical region are points at low $t$. 
Here, following Schwartz \cite{schwartz_blog}, we
perform the expansion in small $t$
\begin{eqnarray}
f(t) &=& f_0 + f_2 t^2 + f_3 t^3 + f_4 t^4 \ln(t) + O(t^4)\,,\nonumber \\
f_3 &=& 8 Z D, \qquad f_4 = 16 Z^2 D\,. 
\label{ftexp}
\end{eqnarray}
The value of the integrand in Eq. (\ref{intlnk0}) at $t=0$ is equal to $f_3$
and this helps to judge the numerical precision achieved at small $t$-region.
In order to perform integration, we use
polynomial interpolation of the integrand for $t>0.1$.
For the remaining region, $t \in [0,0.1]$ we extend the expansion of the remainder denoted by
$O(t^4)$ in Eq. (\ref{ftexp}) to higher powers of $t$, 
next we fit the expansion terms and finally perform the integration.
 
The function $f(t)$ can be represented as the matrix element with $\Psi$ and a pseudo-state 
function $\delta \Psi_{\rm in}$ given by 
\begin{eqnarray}
f(t) &=& -\omega\,\langle \Psi | \vec P | \delta \Psi_{\rm in} \rangle \\
({\cal E}_0-{\cal H}_0-\omega) \, \delta \Psi_{\rm in} &=& \vec P \Psi \,.
\label{eqdpsi}
\end{eqnarray}
In the calculations of the Bethe logarithm for  $S$-states, $\delta \Psi_{\rm in}$ is 
of the form of a P-function in Eq.~\eqref{psiP}. 
This function is calculated by optimizing
$f(t)$ against nonlinear parameters of $\delta \Psi_{\rm in}$ for each $t$ individually. 
The special case is $f(t)/\omega$ in the limit $\omega=0$,
which is established by the Thomas-Reiche-Kuhn sum rule for dipole oscillator strengths 
$\langle \vec P \,({\cal H}_0-{\cal E}_0)^{-1}\,\vec P \rangle = 3/2 \, Z$
\cite{bethe}. 
It helps in judging on the completeness of the intermediates states and on the estimation of uncertainties.

Calculations for the $2 P$ state are more sophisticated as
compared to the ground state. The dipole operator $\vec P$ couples the wave function of the 
$2 P$ state to three classes of intermediate states $S$, $P^e$, and
$D$ which differ by the orbital angular momentum.  
This fact allows to split the optimization of 
$\delta \Psi_{\rm in}$ into three parts, involving $\phi_S$, and 
\begin{eqnarray}
\phi^{i}_{P^e} &=& \epsilon_{ijk} r_a^i r_b^j \,\phi_S\,,\\
\phi^{ij}_{D} &=& \bigg(\frac{r_a^i r_b^j + r_b^i r_a^j}{2} - 
\frac{\delta^{ij}}{3}\,\vec r_a \cdot \vec r_b \bigg)\,\phi_S,
\end{eqnarray}
respectively.
In order to be able to use the optimization against nonlinear parameters,
the negative contribution coming from $2 S$ to $\delta \Psi_{\rm in}$
is represented as a fixed sector with parameters taken from the ground state
wave function. All the other nonlinear parameters are obtained
by optimization of the corresponding parts of $f(t)$.

The mass  coefficient ${\cal E}^{(5,1)}_{\rm mp}$ in Eq.~\eqref{e51}, 
involve the mass polarization correction to the Bethe logarithm. 
The framework of such perturbative calculations have been presented 
on the recoil corrections to the helium atom \cite{helrec}. 
Following the Eq.~\eqref{intlnk0} 
we can build the integral representation in the form
\begin{eqnarray}
\Delta (\ln k_0)_{\rm mp} &=& \frac{1}{D} \int_0^{\infty} dt \,\frac{ f_{\rm mp}(t)-f_{0,\rm mp}-f_{2,\rm mp}\,t^2}{t^3} \nonumber \\
&& - \frac{D_{\rm mp}}{D}\,\ln k_0 
\label{intdlnk0}\\
  f_{\rm mp}(t) &=& -\omega\,\bigg \langle \vec P \frac{1}{{\cal E}_0-{\cal H}_0-\omega} \vec P \bigg \rangle_{\rm mp} 
\label{ftmp}\\
&& \hspace{-1.5cm}  + \omega \,\bigg \langle \vec P \frac{1}{{\cal E}_0-{\cal H}_0-\omega} 
\big(\sum_{a<b} \vec p_a \cdot \vec p_b - \langle \vec p_a \cdot \vec p_b \rangle \big) \\ 
&& \times \frac{1}{{\cal E}_0-{\cal H}_0-\omega} \vec P \bigg \rangle \nonumber \\
D_{\rm mp} &=& 2\, \pi\,Z \,\sum_a  \langle \delta^3(r_a) \rangle_{\rm mp}, \\
f_{0, \rm mp} &=&  \langle  \vec P^2 \rangle_{\rm mp}, \quad f_{2, \rm mp} = -2\,D_{\rm mp}\,,
\end{eqnarray}
Apart from the correction to the wave function Eq.~\eqref{psimp}, 
the mass polarization correction Eq.~\eqref{hmp} 
in the denominator of the $f(t)$ in the Eq.~\eqref{ft} 
corresponds to the second term of Eq.~\eqref{ftmp}. 
All terms in the above representation are evaluated as the mean values with numerically 
determined functions $\Psi$, $\delta \Psi_{\rm mp}$ and $\delta \Psi_{\rm in}$. 
\section{Results}

\begin{table}[!thb]
\renewcommand{\arraystretch}{1.0}
\caption{Bethe logarithm and the mass polarization correction in the lithium atom}
\label{tableBLog}
\begin{ruledtabular}
\begin{tabular}{cw{3.6}w{3.6}w{3.6}}
Shell                                 & \cent{\ln k_0} & \cent{\Delta \ln k_0} \\
\hline  \\
\multicolumn{3}{c}{$2 S$}\\
$7$                          & 5.178\,067\,3 &    0.113\,813   \\
$8$                          & 5.178\,145\,2 &    0.113\,776  \\
$9$                          & 5.178\,161\,8 &    0.113\,835  \\
$10$                         & 5.178\,167\,9 &    0.113\,807	  \\
$\infty$                     & 5.178\,169(4) &    0.113\,81(3)  \\ \\
 Pachucki, Komasa (2003) \cite{ligauss}      & 5.178\,17(3)  &  0.114(3)   \\
 Yan {\it et al.} (2008) \cite{yan_rel}      & 5.178\,28(1)  &  0.113\,05(5)    \\
\\
\multicolumn{3}{c}{$2 P$}\\ 
$7$                          & 5.178\,974   &  0.116\,09  \\
$8$                          & 5.179\,549   &  0.113\,16   \\
$9$                          & 5.179\,635   &  0.111\,96   \\
$10$                         & 5.179\,780   &  0.111\,48   \\
$\infty$                     & 5.179\,81(7) &  0.111\,3(5)          \\\\
Yan {\it et al.} (2008) \cite{yan_rel}  &  5.179\,79(6)  &  0.111\,2(5)    \\
\end{tabular}
\end{ruledtabular}
\end{table}

Table~\ref{tableBLog} presents numerical values of the Bethe logarithms for the lowest doublet
S and P states of lithium. In order to control the numerical uncertainty, 
we performed calculations with several basis sets 
successively increasing the shell parameter (see eg. \cite{lit_wave}). 
Our result for $\ln k_0$ and $\Delta \ln k_0$  agrees well with former ones based on the integral 
representation with ECG functions \cite{ligauss}, but are more accurate. An inconsistency is observed 
with the result obtained in Hylleraas basis by Yan {\em et al.} \cite{yan_rel}. These authors use 
a discrete variational representation of the continuum in terms of pseudo states 
to cover a huge range of distance scales \cite{drake_blog}. 
At the present level of theoretical predictions, the resulting differences of $15$ MHz for transition energies 
$2 P - 2 S$ and $1$ kHz for isotope mass shifts are much smaller compared to the uncertainties 
from estimations of higher order QED corrections. For $2 P$ state our results confirm those by Yan {\em et al.} 
\cite{yan_rel} and both have comparable uncertainties. 

Table \ref{table1} presents results for various spin independent  ${\cal E}^{(m,n)}$
dimensionless coefficients. 
Except for $m \alpha^6$, $m \alpha^7$ and $m \alpha^6 \eta$ 
terms, all uncertainties have been obtained from the analysis of the numerical convergence 
as a function of the different basis size for the wave function, since operator structure 
at given order in $\alpha$ and $\eta$ are included in a complete way. 
Complete result for ${\cal E}^{(6,0)}$ is not yet known, it is estimated
by what we think, the dominating one-loop radiative correction, see Eq. (\ref{e60}). 
The uncertainty of  ${\cal E}^{(6,0)}$ is assumed to be 10\%, what is a dominant source of 
the overall uncertainty. Analogous estimations of ${\cal E}^{(7,0)}$ and ${\cal E}^{(6,1)}$ 
are established at the level of 25\%. 
Next columns of Table \ref{table1} present (centroid) transition energies $\nu_{6}$, $\nu_{7}$,
and the (centroid) isotope shift $\Delta \nu_{67}$. One observes that each
power of $\alpha$ or $\eta$ gives significantly smaller contributions, 
so the expansion in $\alpha$ and $\eta$ is physically well meaningful.

\begin{widetext}

\begin{table}[!hbt]
\renewcommand{\arraystretch}{1.0}
\caption{Spin independent contributions (centroid) to the transition energy $\nu(2 P-2 S)$ for $^6$Li, $^7$Li 
and to the isotope shift.}
\label{table1}
\begin{ruledtabular}
\begin{tabular}{cw{4.12}w{10.4}w{10.4}w{10.7} }
energy                 & \cent{{\cal E}^{(m,n)}}     & \cent{\nu_6({\rm MHz})}  & \cent{\nu_7({\rm MHz})} & \cent{\Delta \nu_{67}({\rm MHz})} \\
\hline \\
 $m \alpha^2 $         & 0.067\,903\,791\,259\,0(16) &  446\,785\,483.5(1) & 446\,785\,483.5(1)  &    \\
 $m \alpha^4$          & 0.267\,612\,1(4)            &        93\,765.1(2) &       93\,765.1(2)  &     \\
 $m \alpha^5$          & -3.469(3)                   &        -8\,850.(8)  &       -8\,870.(8)   &    \\
 $m \alpha^6$          &  -14.4(1.4)                 &           -269.(26) &           -269.(26) &    \\
 $m \alpha^7$          &   217.(54.)                 &             30.(7)  &             30.(7)  &    \\
 $m \alpha^2 \eta $ & 0.123\,007\,926\,0(3)       &       -73\,826.6    &       -63\,293.1    &  -10\,533.510\,5  \\
 $m \alpha^4 \eta$  &  -0.267\,591(2)             &              8.6    &              7.3    &         1.220\,2\\
 $m \alpha^5 \eta$  &   1.134(3)                  &             -0.3    &             -0.2    &        -0.037\,7(1)\\
 $m \alpha^6 \eta$  &  -46.(12)                   &                     &                     &         0.011(3)\\
 $m \alpha^2 \eta^2$& -0.004\,870(4)              &                     &                     &        -0.070\,7     \\
 ${\cal E}^{(4,0)}_{\rm fs}$  &  -1.045 608 \, r^2_{\rm c}  &   -16.0  &        -14.0      & \\
 ${\cal C}_{67}$       &                             &                     &                     &    -2.465\,8      \\
 Total                 &                             &  446\,796\,306.(28) & 446\,806\,840.(28) & -10\,532.388(3)
\end{tabular}
\end{ruledtabular}
\end{table}

\end{widetext}

Table \ref{table2} presents results for contributions to the fine structure
splitting and the corresponding isotope shift. 
The result for mixing of $2 P_{3/2}$ and $2 P_{1/2}$ due to hyperfine 
interaction comes from our former paper \cite{lit_fine}. 
Other contributions can in principle be deduced from 
the individual operators given in Ref. \cite{lit_rel,lit_fine},
but results presented here are recalculated with higher numerical precision,
and they are the most accurate among ones available in the literature (except $m \alpha^2$ \cite{Wang_2012}). 
We note that the present numerical precision for expansion coefficients is
high enough, so that  the uncertainty comes exclusively from 
neglected higher order corrections. These corrections are not expected to be very small,  
therefore in comparison to experimental results by Sansonetti {\it et al.}
\cite{sansonetti2011,brown2012} theoretical predictions for the fine structure
are much less accurate. However, the isotope shift 
in the fine splitting is not so much sensitive to higher order terms,
as they cancel out in the difference. As it was pointed out by Yan {\it et al} \cite{yan_rel},
various experiments and theoretical predictions are all in disagreement
between themselves. Only recently, after correcting for the hyperfine mixing
the theoretical predictions of $-0.544\,7(1)$  start to agree with very recent
experimental value obtained by NIST group $-0.531(24)$ \cite{brown2012}.

\begin{widetext}

\begin{table}[!hbt]
\renewcommand{\arraystretch}{1.0}
\caption{The fine structure in $^6$Li, $^7$Li and the fine structure isotope shift $^{6,7}$Li in MHz.
The Total line includes only numerical uncertainties, and not uncertainties from the unknown $m\,\alpha^6$ correction.
This correction is negligible only for the isotope shift in the fine structure.}
\label{table2}
\begin{ruledtabular}
\begin{tabular}{cw{3.15}w{7.7}w{7.7} w{2.10}}
energy                 & \cent{-3/2\,{\cal F}^{(m,n)}(2 P_{1/2})}   &  \cent{\nu_6(2 P_{3/2}-2 P_{1/2})({\rm MHz})}& 
\cent{\nu_7(2 P_{3/2}-2 P_{1/2})({\rm MHz})}  & \cent{\Delta \nu_{67}(2 P_{3/2}-2 P_{1/2})({\rm MHz})} \\
\hline 
 $m \alpha^4 $        &   0.028\,693\,979(3)          &  10\,053.712\,6(11)   & 10\,053.712\,6(11)   &  \\
 $m \alpha^4 \eta$    &   0.087\,18(2)                &       -2.786\,1(5)    &      -2.388\,6(4)       &  -0.397\,5(6) \\
mixing \cite{lit_fine}&                               &        0.012\,17      &       0.159\,16      &  -0.146\,99\\
 Total                &                               &  10\,050.938\,6(12)   &  10\,051.483\,2(12)  &  -0.544\,7(1) \\ \\
 Experiment \cite{sansonetti2011}    &                &  10\,052.799(22)      &  10\,053.393(21)     &  -0.594(30) \\
 Experiment \cite{brown2012}         &                &  10\,052.779(17)      &  10\,053.310(17)     &  -0.531(24) 
\end{tabular}
\end{ruledtabular}
\end{table}

\end{widetext}

Transition energies for D1 and D2 lines in Table \ref{table3} 
are obtained from the data presented in Tables \ref{table1} and \ref{table2}. 
As for the fine structure, theoretical predictions are much less accurate
in comparison to experimental values \cite{yan_rel,sansonetti2011}
due to unknown complete $m\,\alpha^6$ contribution. 
Nevertheless, our results are in good agreement with previous theoretical 
calculations by Yan and Drake in \cite{yan_rel}. The small difference 
comes mainly from previously mentioned difference in ground state Bethe logarithms in Table \ref{tableBLog}.

\begin{widetext}

\begin{table}[!hbt]
\renewcommand{\arraystretch}{1.0}
\caption{Transition energies for D1 and D2 lines. Comparison of theoretical and experimental values.}
\label{table3}
\begin{ruledtabular}
\begin{tabular}{cw{3.15}w{2.15}w{2.15} w{2.15}}
energy                                & \cent{ 2P_{1/2}-2S_{1/2}({\rm MHz})} & \cent{ 2P_{3/2}-2S_{1/2}({\rm MHz})}\\
\hline \\
\multicolumn{3}{c}{$^6{\rm Li}$}\\
 This work                           &    446\,789\,589.(20)       &   446\,799\,640.(20)  \\
 Experiment \cite{yan_rel}           &            &                 \\
 Experiment \cite{sansonetti2011}    &    446\,789\,596.091(7)     &   446\,799\,648.927(21) \\
 Experiment \cite{brown2012}         &                             &   446\,799\,648.870(15) \\
\multicolumn{3}{c}{$^7{\rm Li}$}\\
 This work                           &    446\,800\,123.(20)       &   446\,810\,175.(20)    \\
 Theory \cite{yan_rel}               &    446\,800\,142.(30)       &   446\,810\,193.(30)     \\
 Experiment \cite{yan_rel}           &    446\,800\,130.61(42)     &   446\,810\,189.18(42)   \\
 Experiment \cite{sansonetti2011}    &    446\,800\,129.853(6)     &   446\,810\,183.289(20)  \\
 Experiment \cite{brown2012}         &                             &   446\,810\,183.163(16)  \\
\end{tabular}
\end{ruledtabular}
\end{table}

\end{widetext}

Table \ref{table4} provides a new determination of the difference in squared charge radii $\delta r^2_{\rm c}$ from 
a new measurement of the isotope shift of the D1 and D2 lines for $^7{\rm Li}$ and
$^6{\rm Li}$ \cite{brown2012}. 
The final results $0.705(3)$ from D1 (the most precise in the literature) and  $0.700(9)$ from D2 lines 
are in a very good agreement. They both agree with the determination
based on $3S-2S$ transition $0.731(22)$ \cite{isLi2s3sfinal}, 
but not so perfectly. The difference is about one $\sigma$. 
It nevertheless stands as a confirmation of atomic spectroscopy approach
to determination of the nuclear charge radii.

\begin{widetext}

\begin{table}[!htb]
\begin{minipage}{16.0cm}
\renewcommand{\arraystretch}{1.3}
\caption{Summary of the isotope shift determination of the $^6$Li charge radii from D1 and D2 lines with respect to $^7$Li, 
$r(^7{\rm Li}) =  2.39(3)$ fm \cite{liradius}, the first uncertainty of $\nu_{\rm the}$ comes
from unknown higher order terms, the second uncertainty is due to the atomic mass. $C_{67} = -2.465\,8$ [MHz fm$^{-2}$]}
\label{table4}
\begin{ruledtabular}
\begin{tabular}{rw{6.8}w{7.12}w{2.7}w{1.7}}
     D-line & \centt{$\nu_{\rm exp}$[MHz] Ref. \cite{brown2012}}  &
      \centt{$\nu_{\rm the}$[MHz]}  &
      \centt{$\delta r^2_{\rm c}$[fm$^2$]}& \centt{$r_{\rm c}$[fm]} \\ \hline
$2P_{1/2}-2S_{1/2}$  & -10\,533.763(9)   &  -10\,532.023\,7(28)(2)  &  0.705(3)  &  2.533(28)\\
$2P_{3/2}-2S_{1/2}$  & -10\,534.293(22)  &  -10\,532.568\,2(28)(2)  &  0.700(9)  &  2.533(28)\\
$3S-2S$ \cite{isLi2s3sfinal}& &                                     &  0.731(22) &  2.538(28)  \\
\end{tabular}
\end{ruledtabular}
\end{minipage}
\end{table}

\end{widetext}

\section{Summary} 
We have calculated lithium D-lines including leading QED effects
with high numerical precision. All results, but Bethe logarithms
are in agreement with previous ones obtained in the literature.
A discrepancy is observed for the Bethe logarithm of ground state,
which needs further verification. Although all these Bethe logarithms are somehow close
to hydrogenic value, there is no any computational approach
which takes advantage of it. One always has to 
represent the intermediate states very accurately and this is
a numerically demanding task. All the other corrections
can be obtained with much higher accuracy at almost no computational costs.
In comparison to experimental D-lines, agreement is observed
but theoretical predictions are anyway much less accurate.
The main reason, which is a principal problem in improving theoretical results
for lithium, is the complete calculation of $m\,\alpha^6$ corrections,
which so far has been performed only for He \cite{hela6}.
This is a challenging, but what we think, a feasible task,
with computational approach based on explicitly correlated basis functions,
such as Hylleraas, Slaters or Gaussians. This correction 
cancels out in the isotope shift in fine structure, and here
theoretical results are in agreement with the most recent result 
of Sansonetti {\em et al} \cite{sansonetti2011,brown2012}. 
An additional results of this work is the determination of 
$^{6,7}$Li nuclear charge radius difference from the measured 
and calculated  isotope shift. Our result is maybe not in perfect,
but in one $\sigma$ agreement with that obtained from $3S-2S$ transitions,
what clearly demonstrates the applicability of atomic spectroscopy methods
for accurate determinations of nuclear charge radii.
We should mention, nevertheless, that for the simplest possible atoms,
the hydrogen and the muonic hydrogen, a significant and unresolved discrepancy is
observed for the proton charge radius \cite{pohl}.

\section*{Acknowledgments}
This work was supported by NCN grants No. 2012/04/A/ST2/00105 (K.P.) and No. 2011/01/B/ST4/00733 (M.P.).
D.K. acknowledges additional support from Nicolaus Copernicus University,
grant No. 1100-CH,

\end{document}